**Adaptive design for identifying maximum tolerated dose early to accelerate dose-finding trial**

Masahiro Kojima[1,2]

[1]Biometrics Department, R&D Division, Kyowa Kirin Co., Ltd., Tokyo, Japan.

[2]Department of Statistical Science, School of Multidisciplinary Sciences, The Graduate University for Advanced Studies, Tokyo, Japan.

**Running title**: Early identification of MTD

**Financial support**: None

**Corresponding author**

Name: Masahiro Kojima

Address: Biometrics Department, R&D Division, Kyowa Kirin Co., Ltd.

Otemachi Financial City Grand Cube, 1-9-2 Otemachi, Chiyoda-ku, Tokyo, 100-004, Japan.

Tel: +81-3-5205-7200

FAX: +81- 3-5205-7182

Email: masahiro.kojima.tk@kyowakirin.com

**150-word statement of translational relevance:**

We proposed a novel adaptive design for identifying MTD early to accelerate dose-finding trials. The early identification is determined adaptively depending on the toxicity data of the trial. The early identification of MTD leads to faster progression to a phase II trial and expansion cohorts to confirm efficacy. We confirmed that the design adapting early identification does not degrade accuracy compared to conventional designs. We confirmed the early identification designs reduced the study duration by about 50% from the model-assisted designs.

# Abstract

**Purpose:** The early identification of MTD in phase I trial leads to faster progression to a phase II trial and expansion cohorts to confirm efficacy.

**Methods**: We proposed a novel adaptive design for identifying MTD early to accelerate dose-finding trials. We applied the early identification design to an actual trial. A simulation study evaluates the performance of the early identification design.

**Results**: In the actual study, we confirmed the MTD could be early identified and the study period was shortened. In the simulation study, the percentage of the correct MTD selection in the early identification Keyboard and early identification BOIN designs was almost same from the non-early identification version. The early identification Keyboard and BOIN designs reduced the study duration by about 50% from the model-assisted designs. In addition, the early identification Keyboard and BOIN designs reduced the study duration by about 20% from the TITE model-assisted designs.

**Conclusion**: We proposed the early identification of MTD maintaining the accuracy to be able to short the study period.

# Introduction

In the field of oncology, the objective of phase I dose-finding trials is to identify the maximum tolerated dose (MTD). To achieve this goal, model-assisted designs, which combine the simplicity of a 3+3 design with the superior performance of a continual reassessment method (CRM), have recently been proposed. Recent studies use the model-assisted designs (Clinical-Trials.gov identifier: NCT04926285, NCT04573140, NCT04390737, NCT05024305, NCT04678921, NCT04511039). Phase I trials have the problem that we cannot usually proceed to the next cohort because the dose for the next cohort cannot be determined until the safety assessment completes. For example, rapid patient enrollment or late-onset of toxicity would slow down the study. Therefore, Yuan et al. (1) and Lin et al. (2) proposed a time-to-event (TITE) model-assisted design which determines the dose for the next cohort and proceeds to the next cohort during the safety assessment for the current cohort. However, the study cannot identify the MTD until the number of patients treated reaches the sample size. Early identification methods for MTD have been proposed (3,4). However, the methods cannot apply to TITE model-assisted designs.

In this paper, we propose a novel early identification method of MTD for the TITE

model-assisted designs. The early identification allows to proceed to an expansion cohort or phase II trial quicker to confirm efficacy. The early identification is determined when the MTD is estimated with sufficient accuracy based on the dose maintenance probability. The early identification method performs on an actual trial. A simulation study evaluates the performance of the early identification method.

## Methods

The TITE model-assisted designs include the modified toxicity probability (mTPI) [5], Keyboard (6), and Bayesian optimal interval (BOIN) (7,8) designs. A feature of these designs is to provide the number of DLTs to determine the dose assignment as shown in Table 1 in advance. The dose-retainment probability for the early identification is calculated using this Table 1. We introduce the dose maintenance probabilities. We assume a phase I dose-finding trial with sample size $N$. The total number of patients treated at the current dose is $n$, the total number of DLTs at the current dose is $n_{DLT}$, the total number of no DLTs patients evaluated at the current dose is $n_{noDLT}$, the no DLT time of pending patients at the current dose is $t_{pend}$, the DLT assessment window is $t$, the no DLT time rate is $n_{pend} = \frac{t_{pend}}{t}$, the estimated number of no DLT patients is $n_e = n_{noDLT} + n_{pend}$, the number of remaining patients is $r$, the number of DLTs of dose escalation decision at $n + r$ patients in the dose-assignment table is $E_{n+r}$, the number of DLTs of dose de-escalation decision at $n + r$ patients in the dose-assignment table is $D_{n+r}$. The number of remaining patients and no observed time of pending patients is $r_{pend}$.

The dose-retainment probability is given by

$$BB(D_{n+r} - 1 - n_{DLT}; r_{pend}, n_{DLT}, n_{DLT} + n_e) - BB(E_{n+r} - n_{DLT}; r_{pend}, n_{DLT}, n_{DLT} + n_e).$$

$BB(a;b,\alpha,\beta)$ is the cumulative beta-binomial distribution function with the number of successes $a$, the number of trials $b$, and the beta shape parameter $\alpha$ and $\beta$. $BB(D_{n+r} - 1 - n_{DLT}; r_{pend}, n_{DLT}, n_{DLT} + n_e)$ refers to the dose not de-escalation probability for $r$ patients using the maximum value for dose not de-escalation $(D_{n+r} - 1 - n_{DLT})$. Because, the probability includes the dose escalation probability, we take the difference with the dose escalation probability $BB(E_{n+r} - n_{DLT}; r_{pend}, n_{DLT}, n_{DLT} + n_e)$. The threshold for early identification of MTD is $t$. If the probability of dose maintenance exceeds $t$, the trial can halt and the MTD is identified. The recommended value of the threshold is 0.4. The rationale of the recommended value is explained by Kojima (4). At the maximum dose, because there is no dose escalation, the early identification is determined by $BB(D_{n+r} - 1 - n_{DLT}; r_{pend}, n_{DLT}, n_{DLT} + n_e)$. At minimum dose, because there is no dose de-escalation, the early identification is determined by $1 - BB(E_{n+r} - n_{DLT}; r_{pend}, n_{DLT}, n_{DLT} + n_e)$. The threshold value at the maximum and minimum doses is twice. Hence, the recommended value is 0.8. If there is no DLT at the current dose, the conditional $n_{DLT}$ and $n$ of the cumulative beta-binomial distribution function are added 0.5. The rationale of adding 0.5 is explained by Kojima (4).

We present a numerical example for the BOIN design with a target DLT level 30%. The DLT

assessment window is three months. The patient enrollment is one patient per month. The sample size is $N = 18$, the total patients treated is 12, $n = 9$ patients received the current dose, the total number of DLTs at the current dose is $n_{DLT} = 3$, the total number of no DLTs patients evaluated at the current dose is $n_{noDLT} = 4$, two patients are pending at the current dose。 We illustrate this example in Figure 1. The pending time for the eleventh patient is two months and the pending time for the twelfth patient is one month. The no DLT time of pending patient at the current dose is $t_p = 3$, the total evaluation time of pending patient at the current dose is $t = 3$, the no DLT time rate is $n_p = \frac{t_p}{t} = \frac{3}{3} = 1.0$, the estimated number of no DLT patients is $n_e = n_{noDLT} + n_p = 4 + 1 = 5$, the number of remaining patients including no DLT time is $r = 6$. From Table 1, $E_{n+r} = E_{15} = 3$ and $D_{n+r} = D_{15} = 6$. The dose not de-escalation probability is $BB(3; 6 + 1.0, 3, 5) = 0.500$ and dose escalation probability is $BB(1; 6 + 1, 3, 5) = 0.096$. Hence, the dose retainment probability is 0.404. The early identification of MTD is determined. We can halt the MTD estimation phase. We apply the early identification to an actual trial. The TITE-model assisted design is a new design and there are no completed studies. Hence, the early identification performs for an actual study with TITE continual reassessment method (CRM) design with similar performance to the TITE model-assisted design.

**TBCRC 024 trial as an illustrative example.** The TBCRC 024 trial (9) was a phase I trial using the time-to-event CRM (TITE-CRM) design for the chest wall and regional lymph nodes in patients with inflammatory or locally recurrent breast cancer after complete surgical resection. The primary objective was to determine the MTD of veliparib in combination with chest wall and nodal radiotherapy. The safety assessment period is 10 weeks (a 70-day time period). The planned four dosages of veliparib were 50 mg, 100 mg, 150 mg, and 200 mg, which were taken orally twice a day. The sample size was 30. The target DLT level was 30%. The number of patients treated and DLTs at each dose were 50 mg ($n = 3$, $n\_DLT = 0$), 100 mg ($n = 6$, $n\_DLT = 2$), 150mg ($n = 12$, $n\_DLT = 2$), and 200 mg ($n = 9$, $n\_DLT = 1$). Although we cannot confirm the DLT status for each cohort from the paper, we assume that the dose was not been reduced after administration of 200 mg because the DLT was only observed once at 200 mg. We assume that the enrollment is one patient per 60-day.

[No DLT in the first cohort] We assume that the safety evaluation of two patients treated completed and the third patient with no DLT has been observed for up to 35 days. The dose-retainment probabilities of the TITE-mTPI, TITE-Keyboard, and TITE-BOIN are 0.93. The probability is over the threshold 0.8. Hence, we identify the MTD. By the early identification, the study period was shortened by 395 days (395 days= 25 days (the third patient's remaining safety assessment duration) + 5 × 60 days (the remaining five patients' safety assessment duration) + 70 days (the last patient's safety assessment duration))

[One DLT in the first cohort] We assume that the safety evaluation of two patients treated completed and one patient occurs a DLT. The third patient with no DLT has been observed for up to 35 days. The all dose-retainment probabilities of the three TITE model-assisted designs are 0.55. We cannot identify the MTD early. For the second cohort, the safety evaluation of two patients treated completed and the third patient with no DLT has been observed for up to 35 days. The dose-retainment probabilities of the TITE-mTPI, TITE-Keyboard, and TITE-BOIN are 0.98. The probability is over the threshold 0.8. Hence, we identify the MTD. By the early identification, the study period was shortened by 215 days. (215 days= 25 days (the sixth patient's remaining safety assessment duration) + 2 × 60 days (the remaining two patients' safety assessment duration) + 70 days (the last patient's safety assessment duration)).

We evaluate the performance of the early identification of MTD via a simulation study.

**Numerical simulation study.** We demonstrate a simulation study to compare early identification TITE mTPI (EI-TITE-mTPI), early identification TITE Keyboard (EI-TITE-Keyboard), and early identification TITE BOIN (EI-TITE-BOIN) designs with mTPI, TITE mTPI, Keyboard, TITE Keyboard, BOIN, and TITE BOIN designs. We imitated simulation setup by Lin et al. (2). We assume that the sample size is 36, the dose level is six. The DLT assessment window is three months. The patient enrollment is two patients per month. The target DLT level is 30%. The number of simulations times is 10,000. The threshold for early identification of MTD is 0.4. For the mTPI and Keyboard designs, the proper dosing interval is $(0.25, 0.35)$. For the BOIN design, the proper dosing interval is $(0.18, 0.42)$. The specifications of the TITE designs are listed in Lin. We prepare a fixed scenario and a randomly set scenario for the true DLT rate of each dose. The simulations perform 10,000 times for each scenario. We prohibit the dose skipping for all designs. To avoid assigning many patients treated to the overly DLT dose, we apply the dose elimination rule. However, the study is not terminated even when the lowest dose has high toxicity rate to confirm early identification of MTD. We evaluated each method using the following criteria.

**Evaluation criteria.**

1. The percentage of correct MTD selection (PCMS)

2. The percentage of early identification of MTD

3. Percent change from non-EI version in average study duration

4. Percent change from non-EI version in average sample size

## Results

**Performance for the selection of the correct MTD.** Figure 2 illustrates the percentage of the correct MTD selection (PCMS) for the six fixed scenarios and two random scenarios. The EI-TITE-Keyboard and EI-TITE-BOIN designs have almost the same PCMS as the non-EI version. The EI-TITE-Keyboard design have at most 2.7% lower PCMS in Scenario 2 and most 2.3% higher PCMS in Scenario 5 compared to the TITE-Keyboard design. The EI-TITE-BOIN design have at most 3.8% lower PCMS in Scenario 2 and most 1.5% higher PCMS in Scenario 5 compared to the TITE-Keyboard design. The PCMSs of EI-TITE-mTPI design are lower than the non-EI versions, most 12.0% lower in scenario 2. We showed the detail results of each scenario in Supplemental Table 3 and 4.

**Percentage of early identification of MTD.** Figure 3 illustrates the percentage of early identification of MTD. The percentages of early identification for EI-TITE-mTPI range from 88.4% to 98.0%, with the average of 94.0% for all scenarios. The percentages of early identification for EI-TITE-Keyboard range from 51.2% to 90.8%, with the average of 69.7% for all scenarios. The percentages of early identification for EI-TITE-Keyboard range from 55.3% to 92.5%, with the average of 73.0% for all scenarios. The EI-TITE-mTPI has the highest percentage of early identification. We confirmed that the EI-TITE-Keyboard and EI-TITE-BOIN designs are able to identify early about 70% on average.

**Percent change from non-EI version in average study duration.** Figure 4 illustrates two bar charts of the percent change from model-assisted designs and TITE model-assisted designs to EI-TITE model-assisted designs in average study duration. For the percent change from the model-assisted designs, the EI-TITE-mTPI design reduces the study duration by 49.4% to 82.9%, with an average reduction of 65.3%. Thus, the study duration is reduced by 31.3 months on average. The EI-TITE-Keyboard design reduces the study duration by 39.0% to 71.8%, with an average reduction of 52.9%. Thus, the study duration is reduced by 25.4 months on average. The EI-TITE-BOIN design reduces the study duration by 39.4% to 72.8%, with an average reduction of 53.9%. Thus, the study duration is reduced by 25.9 months on average. For the percent change from the TITE model-assisted designs, the EI-TITE-mTPI design reduces the study duration by 25.2% to 67.8%, with an average reduction of 42.2%. Thus, the study duration is reduced by 11.7 months on average. The EI-TITE-Keyboard design reduces the study duration by 9.6% to 47.5%, with an average reduction of 21.7%. Thus, the study duration is reduced by 6.0 months on average. The EI-TITE-BOIN design reduces the study duration by 10.9% to 49.4%, with an average reduction of 23.2%.

Thus, the study duration is reduced by 6.4 months on average. We show the summary of percent change from model-assisted designs and TITE model-assisted designs to EI-TITE model-assisted designs in average study duration in Supplemental Table 5. We show the average observed study duration in Supplemental Figure 1.

**Percent change from non-EI version in average sample size.** Figure 5 illustrates the percent change from non-EI versions to EI versions in average sample size. The EI-TITE-mTPI design reduces the number of patients treated by 29.8% to 67.7%, with an average reduction of 46.9%. The EI-TITE-Keyboard design reduces the number of patients treated by 9.4% to 41.6%, with an average reduction of 21.2%. The EI-TITE-BOIN design reduces the number of patients treated by 11.3% to 43.5%, with an average reduction of 23.0%.

## Discussion

We proposed a novel adaptive design for identifying MTD early to accelerate dose-finding trials. The early identification is determined adaptively depending on the toxicity data of the trial. The early identification of MTD leads to faster progression to a phase II trial and expansion cohorts to confirm efficacy. We confirmed that the design adapting early identification does not degrade accuracy compared to conventional designs.

We applied the early identification design to an actual trial (TBCRC 024). The MTD were identified early, and we confirmed that the trial could be shortened by about six months to a year.

The simulation study evaluated the performance of the early identification design. We confirmed that the percentage of the correct MTD selection (PCMS) in the early identification Keyboard and early identification BOIN designs was almost same from the non-early identification version. We found that the early identification mTPI design reduced the PCMS from the non-early identification version by about 10% mTPI. The average percentage of early identification was 94.0% for mTPI and approximately 70% for Keyboard and BOIN designs. The mTPI design had a higher probability of dose maintenance than the other designs because the number of DLTs for which dose maintenance was judged was larger, and thus the early completion rate was higher. On the other hand, the PCMS of mTPI design was low because the early identification was determined even in cases that should not have been determined as early identification. The early identification Keyboard and BOIN designs reduced the study duration by about 50% from the model-assisted designs. A 50% reduction in the simulation refers to a reduction of about two years. In addition, the early identification Keyboard and BOIN designs reduced the study duration by about 20% from the TITE model-assisted designs. A 20% reduction in the simulation refers to a reduction of about half year. The early identification Keyboard and BOIN designs reduced the number of cases by about 20% from the non-early identification version. Shortening the study duration and reducing the number of patients treated allow for more efficient drug development, as patients who were scheduled to be treated in the MTD estimation phase can be enrolled in phase II trials or expanded cohorts earlier.

We confirmed that the performance of the early identification Keyboard and BOIN designs is better. There is little difference in performance between the early identification keyboard and BOIN designs, but we recommend the keyboard design because it has slightly better performance.

**Acknowledgments**: The author thanks Associate Professor Hisashi Noma for his encouragement and helpful suggestions. The author also thanks Keisuke Hanada for his helpful suggestions.

**Author's Contributions**

M. Kojima: Conception and design; development of methodology; acquisition of data (provided animals, acquired and managed patients, provided facilities, etc.): analysis and interpretation of data (e.g., statistical analysis, biostatistics, computational analysis); writing, review, and revision of the manuscript; administrative, technical, and material support (i.e., reporting and organizing data, constructing databases); and study supervision.

**Table 1.** Dose escalation and de-escalation boundaries (TTL=0.3)

| Design | Action | Num of patients treated at current dose | | | | | |
|---|---|---|---|---|---|---|---|
| | | 3 | 6 | 9 | 12 | 15 | 18 |
| mTPI | Escalate if Num of DLTs ≤ | 0 | 1 | 1 | 2 | 2 | 3 |
| | De-escalate if Num of DLTs ≥ | 2 | 3 | 4 | 5 | 7 | 8 |
| Keyboard | Escalate if Num of DLTs ≤ | 0 | 1 | 2 | 2 | 3 | 4 |
| | De-escalate if Num of DLTs ≥ | 2 | 3 | 4 | 5 | 6 | 7 |
| BOIN | Escalate if Num of DLTs ≤ | 0 | 1 | 2 | 2 | 3 | 4 |
| | De-escalate if Num of DLTs ≥ | 2 | 3 | 4 | 5 | 6 | 7 |

**Figure 1.** Example

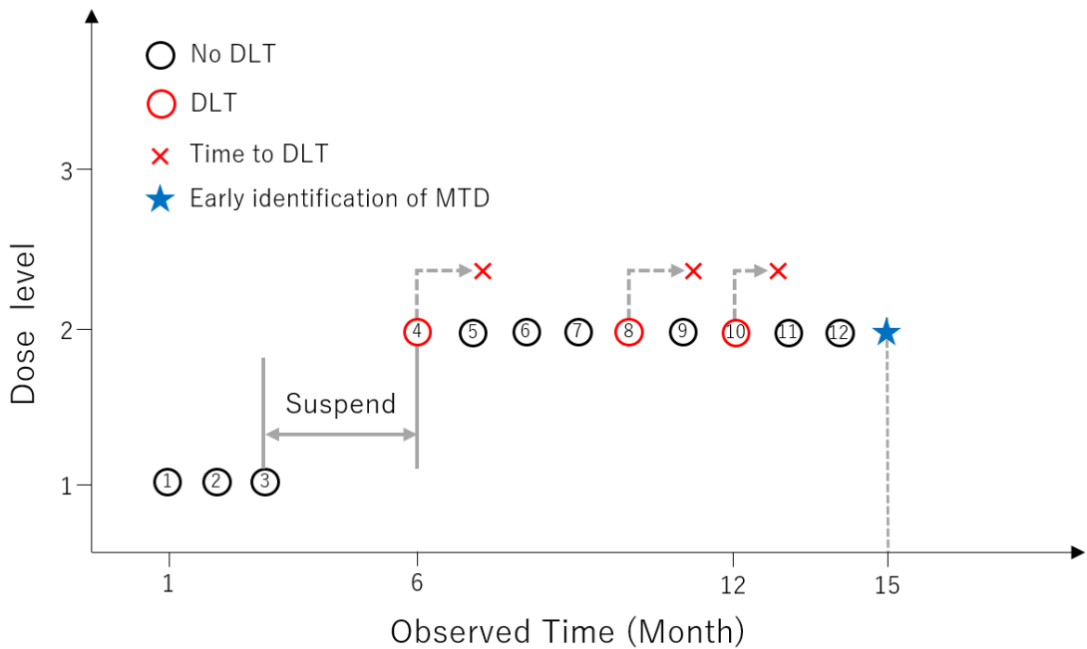

**Figure 2. Percentage of MTD selection**

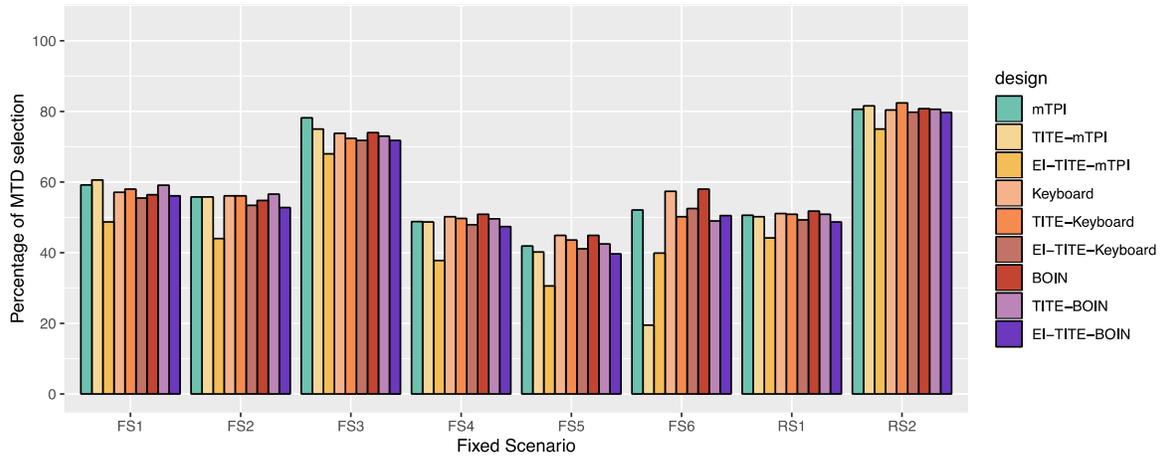

FS1: Fixed Scenario 1; FS2: Fixed Scenario 2; FS3: Fixed Scenario 3; FS4: Fixed Scenario 4; FS5: Fixed Scenario 5; FS6: Fixed Scenario 6; RS1: Random Scenario 1; and RS2: Random Scenario 2. EI: Early identification of MTD

**Figure 3. Percentage of early identification of MTD**

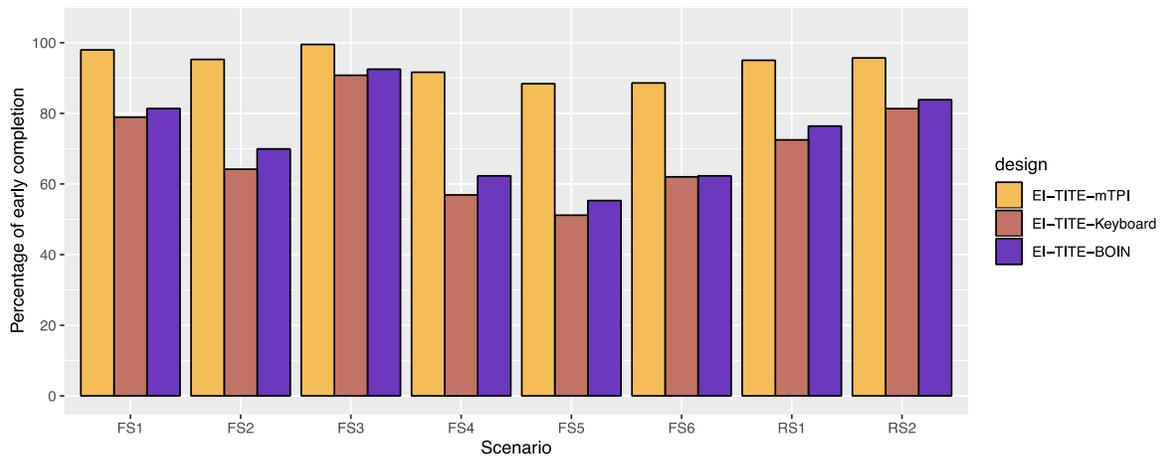

FS1: Fixed Scenario 1; FS2: Fixed Scenario 2; FS3: Fixed Scenario 3; FS4: Fixed Scenario 4; FS5: Fixed Scenario 5; FS6: Fixed Scenario 6; RS1: Random Scenario 1; and RS2: Random Scenario 2. EI: Early identification of MTD

**Figure 4. Percent change non-EI version to EI version in average study duration**
[Percent change from model-assisted designs]

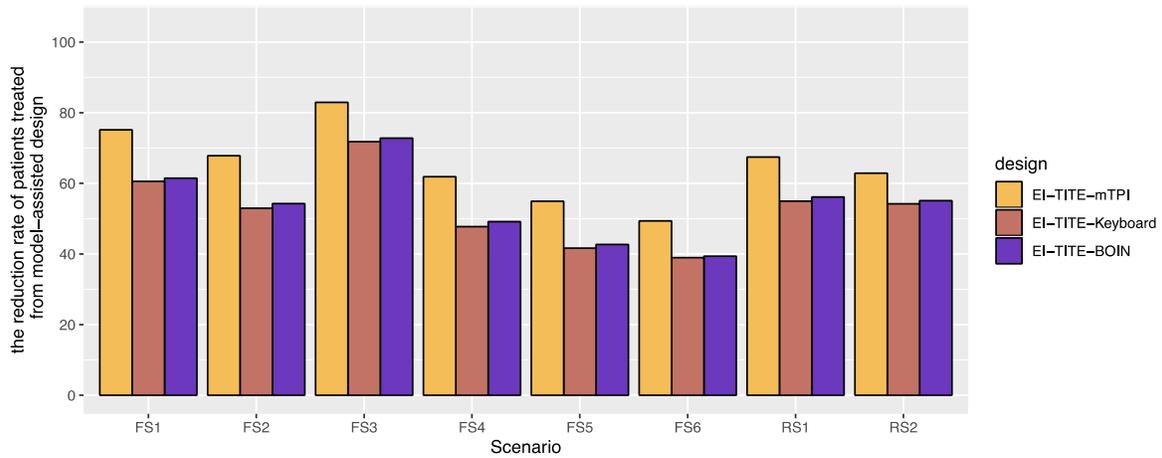

[Percent change from TITE model-assisted designs]

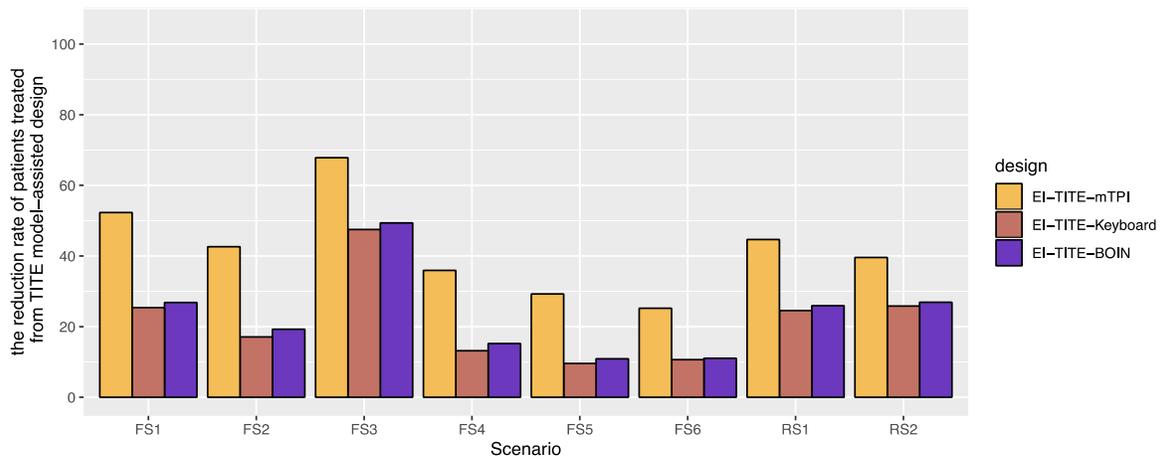

FS1: Fixed Scenario 1; FS2: Fixed Scenario 2; FS3: Fixed Scenario 3; FS4: Fixed Scenario 4; FS5: Fixed Scenario 5; FS6: Fixed Scenario 6; RS1: Random Scenario 1; and RS2: Random Scenario 2. EI: Early identification of MTD

**Figure 5. Percent change from planned sample size in EI version**

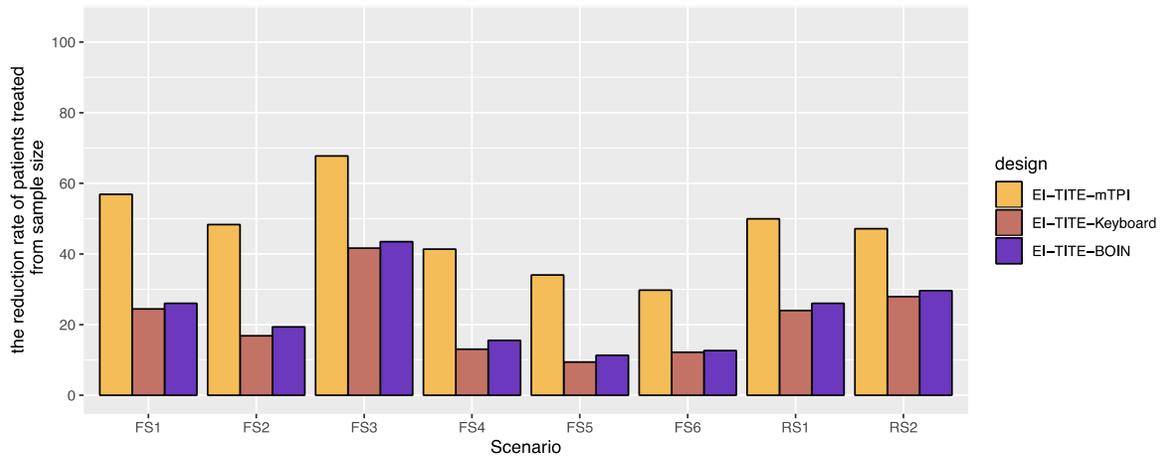

FS1: Fixed Scenario 1; FS2: Fixed Scenario 2; FS3: Fixed Scenario 3; FS4: Fixed Scenario 4; FS5: Fixed Scenario 5; FS6: Fixed Scenario 6; RS1: Random Scenario 1; and RS2: Random Scenario 2. EI: Early identification of MTD